\documentclass[twocolumn,amsmath,amssymb,showpacs,showkeys]{revtex4}

\usepackage{graphicx,color}
\usepackage{dcolumn}
\usepackage{bm}
\newcommand{\beq}{\begin{equation}} \newcommand{\eeq}{\end{equation}}
\newcommand{\bqa}{\begin{eqnarray}} \newcommand{\eqa}{\end{eqnarray}}
 
\newcommand{\erf}[1]{Eq.~(\ref{#1})}

\definecolor{gold}{rgb}{0.75,0.56,0.00}
\definecolor{green}{rgb}{0.00,0.50,0.00}

\begin{document}

\preprint{APS/123-QED}

\title{Rapid state purification protocols for a Cooper pair box}

\author{E. J. Griffith}
 \email{e.j.griffith@liverpool.ac.uk}
\author{C. D. Hill}
\author{J. F. Ralph}
 \affiliation{
 Department of Electrical Engineering and Electronics,\\ The University of Liverpool, Brownlow Hill, Liverpool, L69 3GJ, United Kingdom.}

\author{H. M. Wiseman}
 \affiliation{
Centre for Quantum Computer Technology, Centre for Quantum Dynamics, Griffith University, \\ Brisbane, Queensland 4111, Australia.}

\author{Kurt Jacobs}
 \affiliation{
 Department of Physics, University of Massachusetts at Boston, 100 Morrissey Blvd, Boston, MA 02125, USA.}
 \affiliation{
 Department of Physics and Astronomy, Louisiana State University, 202 Nicholson Hall, Tower Drive, Baton Rouge, LA 70803, USA.}

\date{\today}

\begin{abstract}
We propose techniques for implementing two different rapid state purification schemes, within the constraints present in a superconducting charge qubit system.  Both schemes use a continuous measurement of charge ($z$) measurements, and seek to minimize the time required to purify the conditional state. Our methods are designed to make the purification process relatively insensitive to rotations about the $x$-axis, due to the Josephson tunnelling Hamiltonian. The first proposed method, based on the scheme of Jacobs [Phys. Rev. A {\bf{67}}, 030301(R) (2003)] uses the measurement results to control bias ($z$) pulses so as to rotate the Bloch vector onto the $x$-axis of the Bloch sphere.  The second proposed method, based on the scheme of Wiseman and Ralph [New J. Phys. {\bf{8}}, 90 (2006)] uses a simple feedback protocol which tightly rotates the Bloch vector about an axis almost parallel with the measurement axis. We compare the performance of these and other techniques by a number of different measures.
\end{abstract}



\maketitle


\section{\label{sec:sec1}INTRODUCTION}

Superconducting charge qubits (Cooper pair boxes) are a promising technology for the realisation of quantum computation on a large scale \cite{Nakamura:Nature, You:Scalable}.  For conventional fault-tolerant quantum computing, the quantum states should have a high level of purity, preferably being as close to a pure state as possible.  When the qubit is coupled to an environment  it is subject to decoherence, which will typically result in a completely mixed state \cite{Diosi:Weak}.  However, a qubit initially in a completely mixed state can be purified by measurement. Here we consider continuous measurements, which can be considered as a rapid succession of `weak measurements' \cite{Diosi:Weak}. This gives rise to stochastic `quantum trajectories'  that `unravel' \cite{Car93b} the average density operator evolution described by the Markovian master equation \cite{Scully:Book}.  The quantum trajectory is for the conditional density operator, 
conditioned upon the specific measurement record that was obtained in a given experiment. 

In the Bloch sphere representation, pure qubit states lie on the surface of the sphere, with mixed states being in the interior, and the completely mixed state being at the centre.  When a weak measurement is performed on the qubit, the effect  is to pull the state (on average) towards one of the poles on the measurement axis (which we will take to be the $z$ axis for simplicity).   This pull corresponds to an increase in the average purity over time. An infinitely fast measurement would instantly project the state to one of the poles, as in the quantum Zeno effect \cite{Ruseckas:Zeno}.  However real measurements are never infinitely fast. Moreover, for a charge qubit it can be difficult to connect and disconnect a strongly coupled (fast) measuring device without introducing additional environmental noise. Thus it is necessary to consider measurements giving  a finite rate of purification. 

Since purification takes a finite time, it makes sense to consider whether the information in the measurement record can be used to change the process of purification via feedback. The use of such quantum feedback techniques to increase the purity of conditioned states has been the subject of a number of recent studies \cite{Jacobs:Purity2003,Combes:Purity,Wiseman:Purity}.  Jacobs showed that the maximum increase in the \textit{average purity} occurs when the qubit Bloch vector is rotated onto the $x$-$y$ plane (the plane perpendicular to the measurement axis), after each incremental measurement \cite{Jacobs:Purity2003}. This strategy is optimal in the sense of maximizing the fidelity of the qubit with some fixed pure state at a given final time, as has recently been shown using rigorous techniques from control theory \cite{WisBouHan06}. In addition, this feedback protocol is deterministic because even though the conditioned density operator evolution is stochastic in general, the stochastic term is proportional to the projection of the Bloch vector along the measurement axis \cite{Jacobs:Purity2003}.  Hereafter, in this paper, this protocol is referred to as `ideal protocol \textbf{I}'. 

Although ideal protocol \textbf{I} performs best in the sense just defined, Wiseman and Ralph have recently shown that there are reasons to consider the (conceptually) opposite approach, namely keeping the Bloch vector aligned with the measurement axis \cite{Wiseman:Purity}. They show that this  results in the {\em majority} of qubits reaching a \textit{given} level of purity earlier than in ideal protocol \textbf{I}. In fact this strategy is optimal in the sense of minimizing the \textit{expected time} for a qubit to reach a given level of purity (or fidelity with a fixed pure state) \cite{WisBouHan06}. This is achieved at the expense of having this time be stochastic (unlike ideal protocol \textbf{I}).  Specifically, the distribution of qubit purification times is heavily skewed, with a long tail of low purity values. Hereafter, in this paper, this protocol is referred to as `ideal protocol \textbf{II}'.

This paper addresses a complication which occurs when one attempts to apply either of these two schemes to a specific model of a voltage-controlled charge qubit.  A superconducting charge qubit  consists of a superconducting island (also known as a  Cooper pair box) coupled to a bulk superconductor via a small capacitance and a Josephson weak link junction.  The Josephson junction, which allows the tunneling of Cooper pairs onto and off the island, normally has limited controls.  Although the tunnelling rate can often be modified in experimental systems \cite{Makhlin:Flux}, the Josephson tunnelling energy provides an avoided level crossing between the two qubit energy levels, and maintaining this minimum energy gap minimises the risk of thermal excitation of the system.  Close to this avoided crossing the energy states are formed from superpositions of the quantised charge states (q = 0, 2e) that act as the computational basis for this qubit.  The tunnelling gives rise to a $\sigma_x$ Hamiltonian corresponding to a rotation about the $x$-axis.  As the junction energy should not be zero, the qubit Bloch vector is continually in motion.  The effect of the Hamiltonian evolution often dominates the evolution of the system under the action of the continuous measurements.  This means the Bloch vector can neither be stopped near the $x$-$y$ plane or $z$-axis nor can the direction of rotation be reversed.  However, the applied voltage bias allows some control over the $z$-axis rotations due to the $\sigma_z$ term in the Hamiltonian.  This voltage bias is more commonly expressed as an effective biasing charge $n_g$, \cite{Makhlin:Flux, Gunnarsson:Ext}. 

In this paper we study the purification of charge qubits (Secs.~II and III) using quantum feedback, based on both ideal protocols discussed above.  We suggest mechanisms that could provide near optimal purification rates in the presence of more realistic feedback constraints than those considered in the ideal protcols previously studied, \cite{Jacobs:Purity2003, Wiseman:Purity}. For the protocol based on ideal protocol \textbf{I} we show that good rates for the increase of the  average purity should be achievable using a constant Josephson energy and applying controlled voltage bias field pulses to create a $z$-rotation to rotate the state vector onto the $x$-axis (Secs.~IV and V). We refer to this as practical protocol \textbf{I}.  This is advantageous in two ways: firstly the $x$-axis is trivially on the $x$-$y$ plane which satisfies the rapid purification condition, and secondly the effective radius of the $x$-rotations is reduced, so the vector remains near to the plane even when the control pulses are not accurately applied.  Next, we will show that the Bloch vector can be constrained to the region near the measurement axis ($z$-axis), by using a strong voltage bias field applied at the correct moment to encourage tight radius orbits around an axis almost parallel to the $z$-axis (Secs.~VI and VII). This decreases the average time for the qubit to purify, as in ideal protocol \textbf{II}. We refer to this as practical protocol \textbf{II}. In Section~VIII we conclude with a brief summary of the results. 

\section{\label{sec:sec2}SYSTEM  MODEL}

\begin{figure}[htbp]
	\centering
		\includegraphics[width=0.33\textwidth]{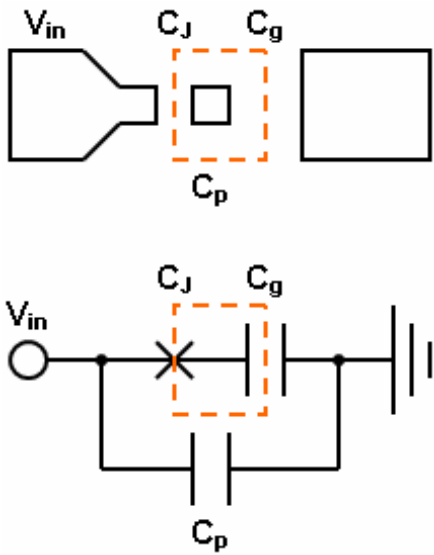}
	\caption{\label{fig:Qubit_model} (Color online) The simple charge qubit model is an island (orange box) coupled to a bulk bias electrode via a single Josephson junction whose frequency is known.  The qubit is controlled via a bias voltage applied across the device.  There are three major capacitances, $C_J$ is the capacitance of the Josephson junction, $C_g$ couples the island to a grounded electrode, and $C_p$ is the parasitic capacitance observed between the bulk electrodes.}
\end{figure}

A superconducting charge qubit, shown in Fig.~\ref{fig:Qubit_model}, consists of a small island of superconducting material connected via a Josephson junction of tunneling frequency $\nu$ to a bulk superconducting electrode, where $\nu=I_C/2$.  The electrode supplies a voltage bias, which can be expressed as an effective charge $q_g$.  The island is also capacitively coupled to a grounded electrode to supply a common reference.  For simplicity, we ignore the dynamics of the biasing circuitry \cite{Griffith:Spec}.

The Hamiltonian of this system is
\begin{eqnarray}
	H & = & \frac{(q-q_g)^2}{2C_q} - \hbar\nu\sin\left(2\pi\frac{\theta}{\Phi_0}\right).
	\label{eq:H}
\end{eqnarray}

\noindent Here $\theta$ is the superconducting phase difference across the Josephson junction expressed in units of the flux quantum $\Phi_0=h/2e$.  There exists a commutation relation between the conjugate variables of charge and phase, $\left[q,\theta\right]=-i\hbar$.  The capacitance $C_q$ is the effective qubit capacitance calculated from the three physical capacitances~\cite{Burkard:Circuit}, $C_J$, $C_g$ and $C_p$, \begin{eqnarray}
	C_q & = & \frac{C_gC_J + C_JC_p + C_pC_g}{C_g + C_p}  
	\label{eq:C_q}
\end{eqnarray}

\begin{figure}[htbp]
	\centering
		\includegraphics[width=0.5\textwidth]{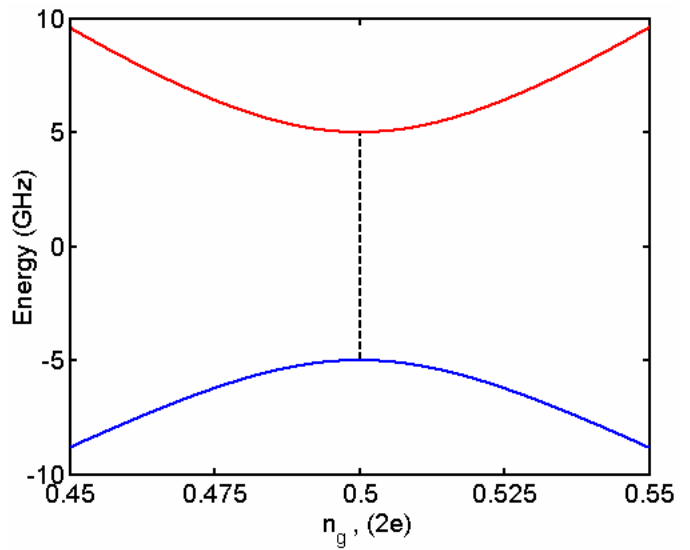}
	\caption{\label{fig:Energy} (Color online) Energy level structure of the qubit.  The avoided crossing caused by the 10GHz Josephson junction creates an energy gap between the energy states.  This separation is kept relatively large to reduce problems caused by thermal excitations or other extraneous noise.}
\end{figure}

\noindent (See Appendix~\ref{sec:appA} for values).  At low energies this Hamiltonian can be approximated by using just two states. Using $n_g=q_g/2e$ for the effective number of Cooper pairs induced by the bias voltage, the Hamiltonian in the charge basis is \cite{Makhlin:Flux, Gunnarsson:Ext}.
\begin{eqnarray}
	\label{eq:Hsigma}
	{H} & = &  \frac{(2e)^2}{2C_q}\left(n_g^2 - n_g + \frac{1}{2}\right)~I \nonumber \\ 
					&   &  + \frac{(2e)^2}{2C_q} \left(\frac{1}{2} - n_g\right)\sigma_z  - \frac{\hbar\nu}{2}~\sigma_x 
\end{eqnarray}
The first term may be discarded as the identity matrix does not affect the dynamics of the system, but is included initially for comparison with \erf{eq:H}.
The second ($\sigma_z$) term shows that the applied voltage bias field controls the rotations about the $z$-axis. When $n_g = 0.5$ the rotations are halted; when $n_g > 0.5$ the qubit rotates in one direction, and when $n_g < 0.5$ the direction is reversed.  The third and final ($\sigma_x$) term is caused by the Josephson junction.  For a single Josephson junction at constant temperature and magnetic fields, the frequency of rotation around the $x$-axis is fixed by manufacture --- we take the frequency to be 10GHz, in line with experimental values \cite{Pashkin:Nature}.  This frequency is equal to the minimum splitting ($n_g$ = 0.5) shown by Figure~\ref{fig:Energy} and it is vital that a sizeable separation is maintained to preserve the two distinct states and suppress the effect of thermal fluctuations.  In some implementations it is possible to use a flux-controlled Josephson junction to vary $\nu$ \cite{Ralph:Aero}. However the method proposed in this paper uses the voltage bias alone to apply the feedback, thereby simplifying the control system.

\section{\label{sec:sec3}CONTINUOUS MEASUREMENT}

For scalable quantum computing \cite{NieChu00} it is necessary to work with pure states.  For qubits, this means states on the surface of the Bloch sphere.  Both pure states (surface of sphere) and mixed states (inside the sphere), can be written as a $2\times2$ density matrix  ${\rho}$.  The on-diagonal elements represent the populations of, and the off-diagonal elements represent the coherences between, the charge states. 
The \textbf{impurity} of the qubit state can be quantified by the following \cite{Combes:Purity}:
\begin{eqnarray}
	L = 1 - P = 1 - {\rm Tr}\left[ {\rho}^2 \right].
	\label{eq:Impurity}
\end{eqnarray}
\noindent Any \textit{pure} state has an impurity of zero, and the \textit{completely mixed} state has an impurity of 0.5.

One way to increase the purity of a state is through measurement, that is to extract some information about the state of the system.  Assuming continuous measurement of charge ($\sigma_z$), the conditioned state $\rho_c$ of the charge qubit obeys a stochastic master equation \cite{WisMil93c} with the following form 
\begin{eqnarray}
		d{\rho_c} & = &  -\frac{i}{\hbar} \left[ {H}, {\rho_c} \right] dt   -\frac{\gamma}{\hbar} \left[ {\sigma_z}, \left[ {\sigma_z}, {\rho_c} \right] \right] dt \nonumber\\ 
					&   & + \sqrt{\frac{2\gamma}{\hbar}} \left( {\sigma_z}{\rho_c} +{\rho_c}{\sigma_z} - 2 \left< {\sigma_z} \right> {\rho_c} \right) dW.
	\label{eq:WeakRho}
\end{eqnarray}
The first term is  the Hamiltonian evolution, with $H$ given by \erf{eq:Hsigma}. The second term represents the back-action of the measurement (parametrized by $\gamma$).  In the absence of Hamiltonian evolution, this causes a deterministic decay of the mixed state towards the $z$-axis.  (We have assumed that there are no other sources of decoherence for simplicity). The third term is due to conditioning upon the measurement result. It is stochastic, with $dW$ being a Wiener increment  \cite{Gar85,Diosi:Weak}. That is, in every time interval of duration $dt$, $dW$ is an independent Gaussian distributed random variable, with  zero mean and variance equal to $dt$. This stochastic term depends upon the particular unravelling considered \cite{Car93b,WisMil93c,Wiseman:Jumps}, which depends upon the measurement scheme. In this case we are conditioning the state upon a continuous `current' which is different in every run of the experiment and which is given by
\beq
I(t)dt = \frac{4\gamma}{\hbar} {\rm Tr}[\rho_c(t) \sigma_z]dt + \sqrt{\frac{2\gamma}{\hbar}}dW(t)
\eeq
This is a current in the generalised sense used in quantum optics and other areas, such that $I(t)dt$ is dimensionless. If this measurement result is ignored then one simply averages over the last term in the stochastic master equation (\ref{eq:WeakRho}). This yields the deterministic master equation given by the first two terms of \erf{eq:WeakRho}. 

Rather than evolve equation (\ref{eq:WeakRho}), we implemented the simulations using Bloch coordinates $v=[x,y,z]^T$ which is equivalent to using the density matrix formalism, however the positional coordinates are easier to visualise.  The equations for the incremental changes in the Bloch coordinates due to continuous weak measurement $dv=[dx,dy,dz]^T$ can be found in Appendix \ref{sec:appB}, in addition the rotations due to the non-zero Hamiltonian acting on the Bloch vector must be included. The equations for $dx$, $dy$ and $dz$ are then numerically integrated over time.

As the system evolves under continuous measurement, it tends to be pulled towards the surface of the Bloch sphere as information (the measurement record) is obtained.  In the absence of Hamiltonian evolution this Bloch vector will be aligned with the measurement ($z$) axis, and the system will evolve stochastically towards one or other of the two poles defined by this axis through the Bloch sphere.  If the Hamiltonian evolution is included and if $\gamma$ is relatively small, the Bloch vector will rotate under the action of the Hamiltonian and be only weakly perturbed by the measurement interaction.  The information extracted by the measurement is dependent upon the orientation of the Bloch vector with respect to the measurement axis \cite{Jacobs:Purity2003}. This means that manipulating  the Hamiltonian by external controls can affect the way that the purity of the system increases.  It is this that forms the basis of the rapid purification protocols discussed in the following four sections.


\section{\label{sec:sec4}FEEDBACK  PROTOCOLS I} 

In this and the following section we are concerned with protocols for maximizing the increase in the \textit{average purity}, as in ideal protocol {\bf I}.  Thus we need a baseline by which to compare the various methods. This baseline is given by ideal protocol {\bf II} --- a situation in which the ideal feedback controls cancel any Hamiltonian evolution and the qubit Bloch vector is allowed to drift stochastically towards the poles.  This appears to be the worst in terms of  the time $T(\epsilon)$ taken for the average impurity $\bar{L}$ to drop to a given level $\epsilon$. For the ideal protocol {\bf II} this function $T_{\bf II}(\epsilon)$ 
is defined implicitly by $\bar{L}_{\bf II}(T_{\bf II}) = \epsilon$, where \cite{Jacobs:Purity2003,Wiseman:Purity}
\begin{eqnarray}
	\bar L_{\bf II}(t) & = & \frac{e^{-4\gamma t}}{\sqrt{8\pi t}} \int_{-\infty}^{+\infty} \frac{ e^{-x^2/\left(2t\right)} }{\cosh\left(\sqrt{8\gamma}x\right)}~dx
	\label{eq:purity_worse}
\end{eqnarray}
This integral can be solved numerically \cite{Jacobs:Purity2003}, but for long times (small $\epsilon$) an analytical approximation gives $T_{\bf II} \sim \ln(\epsilon^{-1})/4\gamma$ \cite{Jacobs:Purity2003,Wiseman:Purity}. For shorter times (larger $\epsilon$) $T_{\bf II}$ is bigger than this expression \cite{Jacobs:Purity2003}. In general, $T_{\bf II}$ can be used to define the speed up factor for a given test method,
\begin{eqnarray}
		S_{\rm test}(\epsilon) & = & \frac{T_{\bf II}(\epsilon)}{T_{\rm test}(\epsilon)}		.
	\label{eq:SpeedUp}
\end{eqnarray}

\subsection{\label{sec:sec4pA} Ideal Protocol I}

The ideal protocol \textbf{I} \cite{Jacobs:Purity2003} rotates the Bloch vector onto the plane orthogonal to the measurement axis to maximise the increase in the \textit{average purity}  during each incremental step.  In our situation, this means rotating onto the $x$-$y$ plane. This protocol eliminates  the stochastic contribution to the evolution of $\rho_c$, as can be verified from the stochastic equations in Appendix~\ref{sec:appB}. Thus the impurity equals the average impurity, which decays exponentially:
\beq \label{eq:PurityBest}
\bar{L}_{\bf I}(t) = e^{-8\gamma t}\frac{1}{2}
\eeq
From this, the time taken to reach impurity $\epsilon$ is
\beq
T_{\bf I} = \ln(\epsilon^{-1}/2)/8\gamma,
\eeq
and for minimizing this time this an exceptional protocol. Thus the maximum speed up factor is 2, in the limit of very small $\epsilon$. 

It would be extremely difficult to apply these instant and perfect control fields to a practical qubit.  In addition, for our superconducting charge qubit there is also the continual motion of the state vector due to the non-zero Josephson junction energy.  The protocols discussed below address these issues. 

\subsection{\label{sec:sec4pB} Flux-controlled Hamiltonian Feedback}

\begin{figure}[htbp]
	\centering
		\includegraphics[width=0.5\textwidth]{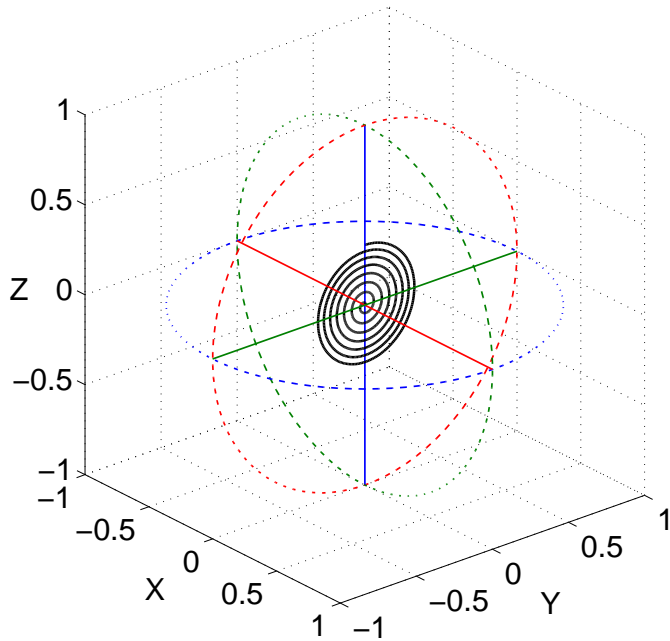}
	\caption{\label{fig:Spiral} (Color online) Under continual $x$-rotation and the influence of weak measurement, the time evolution of the Bloch vector takes a spiral path.  The weak measurements increases the purity, as the purity is a direct measure of the Bloch vector length, the length of the vector is increasing.  As there is only one axis of rotation, the vector length corresponds to the radius of the orbit.  Hence the spiral path found in the $yz$-plane only. 
	\\
	(Note that this should not be interpreted as the average position of the Bloch vector, as the stochastic measurement noise causes random initial phases for the rotation, and as such the average position of the Bloch vector is the centre of the Bloch sphere for all time. This is also true for Figures \ref{fig:ClampX}, \ref{fig:Timing} and \ref{fig:ClampZ}, which are only provided to illustrate the feedback concepts.)}
\end{figure}

Although this is not feedback, natural $x$-axis rotations take the Bloch vector through the $x$-$y$ plane (Fig.~\ref{fig:Spiral}), so there is still an improvement over having no Hamiltonian evolution at all.  The spiral path only momentarily passes through the $x$-$y$ plane so it does not experience the full benefit of the rapid purification protocol.  However, it is possible to utilise a method which uses a flux-controlled Josephson junction to manipulate the $x$-axis rotational frequency \cite{Ralph:Aero}.  The algorithm slows the qubit whilst the Bloch vector is close to the $x$-$y$ plane and then hastens the passage through the $z$-axis poles, maximising the time close to the $x$-$y$ plane.  The control is purely via the $\sigma_x$ term (modulating the Josephson tunnelling frequency, $\nu$) and always maintains a significant energy gap to suppress the effects of thermal fluctuations.  This benefit of this approach is significant \cite{Ralph:Aero} but not as close to ideal protocol {\bf I} as the following approach.  It is important to note that Figure~\ref{fig:Spiral} should not be interpreted as the average position of the Bloch vector, as the stochastic measurement noise causes random initial phases for the rotation, and as such the average position of the Bloch vector is the centre of the Bloch sphere for all time. This is also true for Figures \ref{fig:ClampX}, \ref{fig:Timing} and \ref{fig:ClampZ}, which are only provided to illustrate the feedback concepts.

\subsection{\label{sec:sec4pC} Practical Protocol I}

\begin{figure}[htbp]
	\centering
		\includegraphics[clip,width=0.4\textwidth]{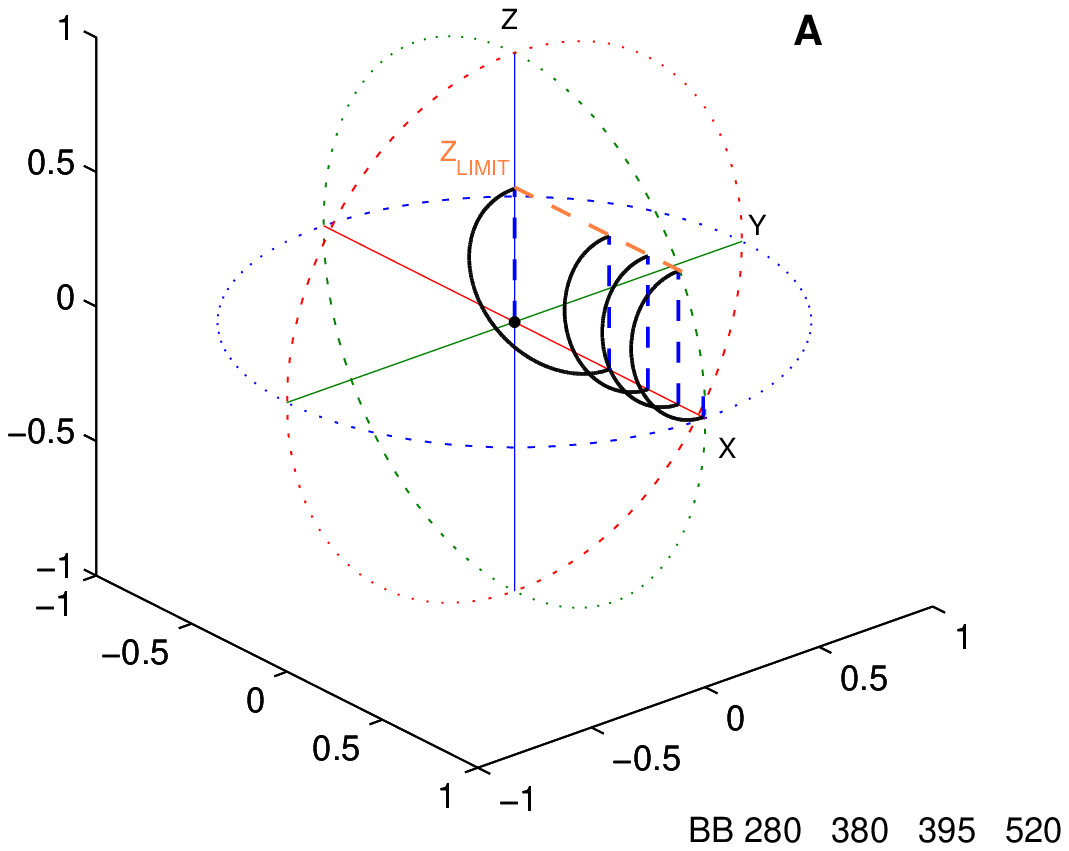}
		\includegraphics[clip,width=0.4\textwidth]{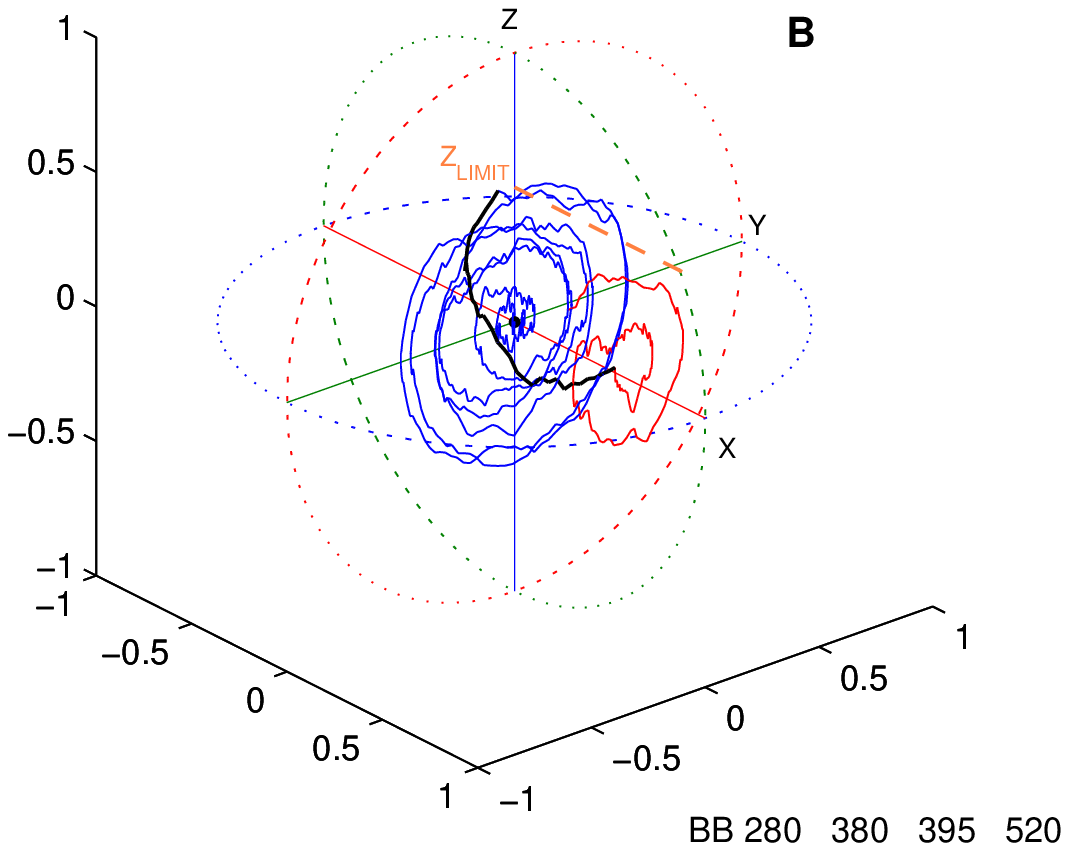}
	\caption{\label{fig:ClampX} (Color online) (\textbf{A}) Pictorial representation of the proposed practical protocol \textbf{I} which attempts to return the Bloch vector to lie on the $x$-axis.  This is achieved using pulsed rotations around the $z$-axis, the combined rotational frequencies of the $x$ and $z$ rotations define a plane of rotation about an arbitrary axis.  The timing and duration is selected to move the vector from the top of the circular path to the bottom ($x$-axis).  This quick corrective feedback is performed whenever the weak measurement process pulls the vector past a predetermined threshold (here $z_{\rm Limit} = 0.5$), creating a screw-like path.  The dashed vertical lines represent the gradual spiralling observed in Figure~\ref{fig:Spiral}. (\textbf{B}) An example trajectory showing the effects of the stochastic continuous measurement noise for the first application of the feedback pulse train.  The noise corrupted feedback path finished above the target $x$-axis, where the second spiral is highlighted in red.}
\end{figure}

\begin{figure}[htbp]
	\centering
		\includegraphics[width=0.5\textwidth]{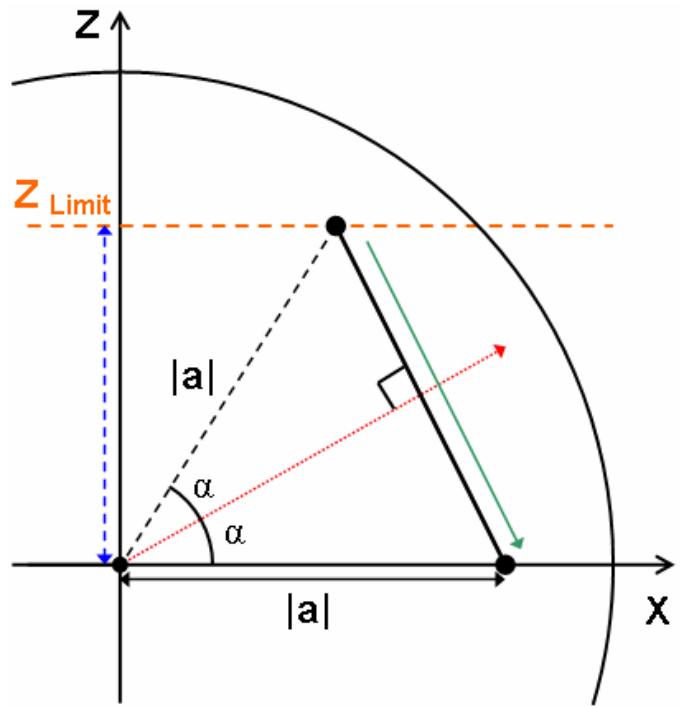}
	\caption{\label{fig:Geometry} (Color online) The applied feedback creates a rotation about the axis (dotted arrow) defined by $\alpha$, this angle can be calculated through a measurement, or estimate, of $z$ and either the purity, $P$ or impurity, $L$.  The $z$ rotation frequency is adjusted and pulsed to ideally take the Bloch vector to ($|\rm{a}|$,0,0) within a half cycle.}
\end{figure}

The first novel algorithm proposed in this paper attempts to use finite duration voltage bias pulses to rotate the Bloch vector repeatedly on to the $x$-axis, taking a screw-like path (Fig.~\ref{fig:ClampX}A), we refer to this as practical protocol \textbf{I}.  The $x$-axis is of particular interest as it is invariant under $x$-rotation. Therefore if the vector can somehow be positioned close to the $x$-axis, it should remain close to the $x$-$y$ plane, even in the absence of further control pulses.  This is why the $x$-axis is an attractive target in the presence of continuous $x$-rotation due to Josephson tunnelling.  However, the effect of the weak measurement is to pull the Bloch vector towards the poles, so the Bloch vector will be gradually pulled away from the $x$-axis in a growing spiral path.  Therefore, to successfully return the Bloch vector to the $x$-axis, a simple control scheme has been devised which utilises a finite duration Hamiltonian proportional to $\sigma_z$ to return the vector to the $x$-axis within a half cycle.  The feedback process triggers when $z$ exceeds a particular threshold $z_{\rm Limit}$ (Fig.~\ref{fig:Geometry}).  On triggering, the controller applies a bias field ($z$-axis rotation) of the required amplitude and duration. An advantage of this approach is that the control field does not need to be continually altered.

After the Bloch vector has reached the $x$-axis, the bias field is removed, so the qubit only experiences the constant $x$-rotation. This creates a distorted spiral path (similar to that in Fig.~\ref{fig:Spiral}) which will once again expand to exceed $z_{\rm Limit}$, where the feedback will trigger.  The overall effect is to constrain the magnitude of $z$ to $z_{\rm Limit}$, so that the Bloch vector remains relatively near the $x$-axis. 

Ignoring the effects of weak measurement during the feedback pulse, it is possible to determine the pulse amplitude and duration analytically. Consider a point in the $xz$-plane with $z=z_{\rm Limit}$ (Fig.~\ref{fig:Geometry}) and use a $\pi$-rotation about an axis (dotted arrow) to finish exactly on the $x$-axis, by rotating from the very top to the very bottom of the circular path about this axis.  The angle $\alpha$ of the axis can be calculated using elementary geometry from $z_{\rm Limit}$-axis and the distance from the centre of the Bloch sphere ($|\rm{a}|=1-2L$). The latter can be obtained from the conditional density matrix $\rho_c$ (which represents the current state of knowledge of the system). We find 
\begin{eqnarray}
	\alpha & = & \frac{1}{2}\sin^{-1}\left(\frac{z_{\rm Limit}}{|\rm{a}|}\right)
	\label{eq:alpha}
\end{eqnarray}
This angle is bounded as follows:
\begin{subequations}	
\begin{eqnarray}
	\alpha_{\rm max} & = & \frac{1}{2}\sin^{-1}\left(1\right) = 45^\circ ~~~~~(x=0)
	\label{eq:maxalpha}
	\\
	\alpha_{\rm min} & = & \frac{1}{2}\sin^{-1}\left(z_{\rm Limit}\right)~~~~~~(|\rm{a}|=1)
	\label{eq:minalpha}
\end{eqnarray}
\end{subequations}

To implement this control strategy it is necessary to determine the $x$ and $z$ angular velocity components. To simplify matters we assume that the  Josephson junction angular frequency is fixed,  $\omega_x = \nu$.  Then we find we require
\begin{eqnarray}
	\omega_z & = & \omega_x\tan\alpha
	\label{eq:Fz}
\end{eqnarray}
 the required bias value can then be obtained from equation (\ref{eq:Fz_ng}) below. Figure~\ref{fig:Timing} shows the relation between the measured $z$ value and the application of feedback as a function of time for the first two applications of the feedback protocol. For this example $z_{\rm Limit} = 0.333$ and the other values can be found in Appendix~\ref{sec:appA}.  The bias pulse duration $\tau$, required to perform the $\pi$-rotation about the $\alpha$ axis is
\beq \label{eq:tau}
\tau = \frac{1}{2}\frac{2\pi}{\sqrt{{\omega_x}^2 + {\omega_z}^2}}
\eeq

\begin{figure}[htbp]
	\centering
		\includegraphics[width=0.5\textwidth]{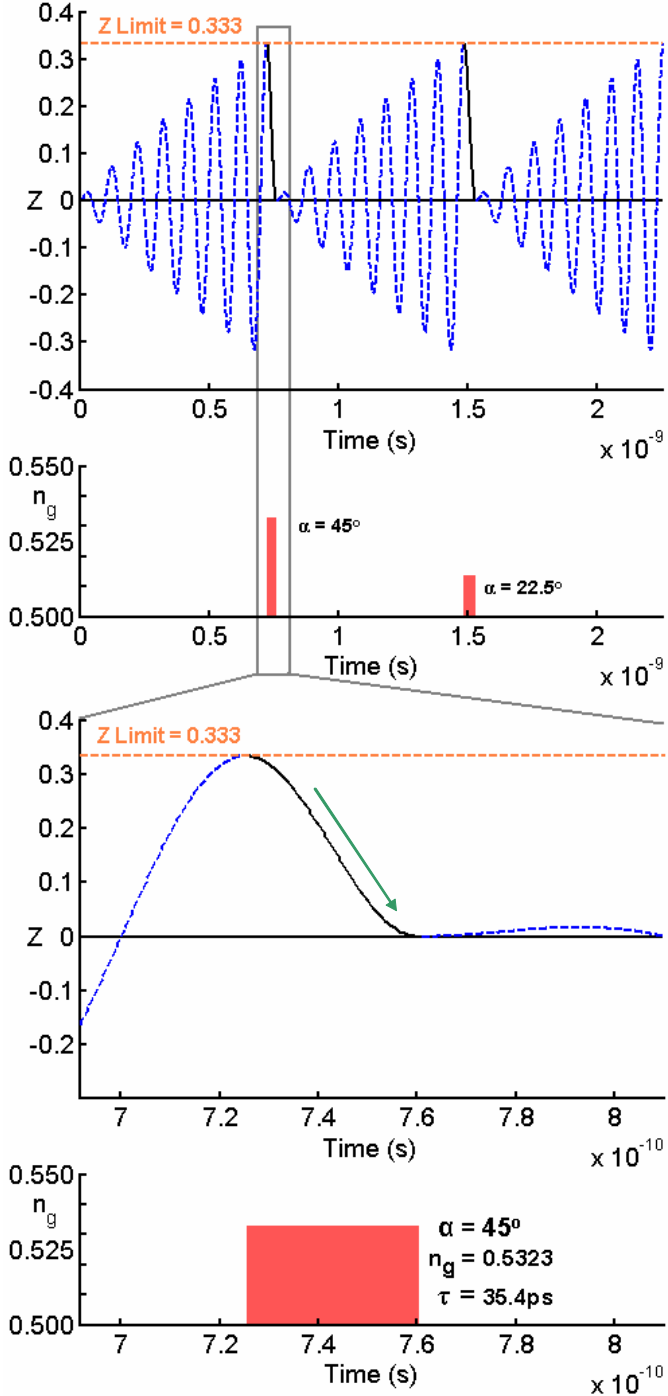}
	\caption{\label{fig:Timing} (Color online) Idealised timing diagram for the feedback pulses with $z_{\rm Limit} = 0.333$ and using the values found in appendix~\ref{sec:appA}.  The topmost graph shows the ideal evolution of $z$ as a function of time, with the dashed and solid line segments corresponding to those found in Figure~\ref{fig:ClampX}A.  The dashed lines indicates no feedback ($n_g = 0.5$), and so the sinusoids are rotations about the $x$-axis due to the tunnelling across the Josephson junction (Fig.~\ref{fig:Spiral}).  When $z_{\rm Limit} = 0.333$ is exceeded, the feedback is applied, and the Bloch vector is $\pi$-rotated from the top of the tilted plane to the bottom such that $z = 0$.  The process is repeated when $z$ again exceeds $z_{\rm Limit}$, the pulses eventually tending to the same shape set by $\alpha_{\rm min}$ (Eqs.~\ref{eq:minalpha},\ref{eq:tau},\ref{eq:Fz_ng})}
\end{figure}

The relation between the $z$-axis rotational frequency and bias is a linear relationship described by the following equation:
\begin{eqnarray}
	\omega_z  & = & \frac{(2e)^2}{\hbar C_q} \left( \frac{1}{2} - n_g \right)
	\label{eq:Fz_ng}
\end{eqnarray}
For our feedback mechanism presented we find that the maximum $\omega_z$ is equal to the Josephson junction frequency, which means there is a limit on the size of bias which should be applied.  This is actually a favourable constraint as applying a bias field substantially larger than $n_g > 0.75$ increases the risk of accessing an unwanted third charge state \cite{Gunnarsson:Ext, Yamamoto:Nature}.  Another requirement of the control system is being able to halt the $z$-rotations or at least slow the rotations significantly by setting the $n_g$ close to 0.5. It is expected that both of these requirements should be achievable. The bias control range to compensate for a system with a constant 10 GHz $x$-axis rotation and the capacitances provided in Appendix~\ref{sec:appA} is:
\begin{eqnarray}
		0.5000 \leqslant n_g \leqslant 0.5323.
	\label{eq:Bias_range}
\end{eqnarray}

The pulse train featured in Figure~\ref{fig:Timing} shows a decreasing trend in the voltage bias amplitude ($n_g$) and an increase in the pulse duration ($\tau$). This increase is due to the slower $z$-axis angular velocity at the latter stages.  The values of $n_g$ and $\tau$ tend to steady state values defined by $\alpha_{\rm min}$, (\erf{eq:minalpha}). Of course, as Figure~\ref{fig:ClampX}B shows, the evolution of the Bloch vector will not be as smooth or predictable as that depicted in the illustration Figure~\ref{fig:ClampX}A, in addition the control pulse cannot be applied or removed instantaneously. There will always be some control delay and some uncertainty as to when the Bloch vector is likely to exceed the threshold value. The stochastic terms will make the prediction of the evolution uncertain and, consequently, the timing of the pulses will contain an uncertainty. To mitigate potential problems in the timing of the control pulses, the evolution of the Bloch vector can be filtered further to reduce the effect of the noise (the evolution described by equation (5) already represents a filter of the information extracted from the qubit). However, numerical calculations including timing errors show that the accuracy of the control pulses is not critial to the performance of the purification protocol (see Figure 10 below). As long as the Bloch vector is reasonably close to the threshold value and the pulse takes the Bloch vector back to the vicinity of the $x$-axis, then the majority of the available purification increase is still obtained. This is good news for a practical implementation of this protocol because it demonstrates that the approach is robust to errors in the control system.

\section{\label{sec:sec5}RESULTS I}

As mentioned in the previous section, finding the peak value of $z$ could prove difficult under noisy conditions.  However, we have found that finding the peak is not required if the measurement of the system is sufficiently weak such that the threshold $z_{\rm Limit}$ is not far exceeded.  Using, a simple threshold on the $z$ value derived from $\rho_c$, the resulting performance is still close to the ideal.  

\begin{figure}[htbp]
	\centering
		\includegraphics[width=0.5\textwidth]{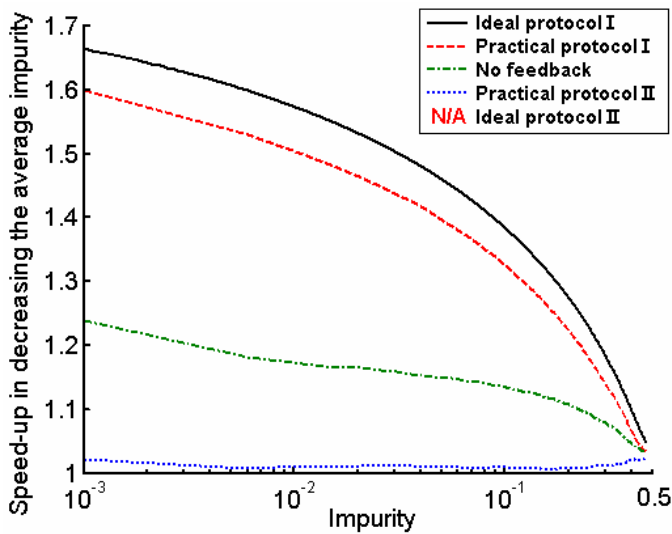}
	\caption{\label{fig:SpeedUp} (Color online) The graph of purity rate improvement shows that the practical feedback protocol proposed in this section (dashed line) achieves similar performance to that of the ideal method (solid line), which required instantaneous feedback.  In addition, the feedback mechanism described in section~\ref{sec:sec6pC} (dotted line) has little performance gain, as expected.  It should also be noted that using no feedback in this system still yields a minor improvement (dash-dot line) over having no Hamiltonian evolution at all}
\end{figure}

Figure~\ref{fig:SpeedUp} shows the improvement in the average purification rate, or (for convenience) the `speed-up', \cite{Jacobs:Purity2003}.  This improvement is measured relative to the minimum \textit{average purification} protocol, the case of free measurement evolution with no Hamiltonian (\erf{eq:purity_worse}), which forms the baseline performance.  The shape of the graph indicates that the performance increase is not constant for all values of purity, with maximum gains at high purity (the final part of the time evolution).  It should be noted that this graph is also independent of the measurement strength $\gamma$.  Equation (\ref{eq:SpeedUp}) defines the improvement $S$ as a function of remaining impurity $L$, and exact analytical solutions this equation are non-trivial, however it is possible to invert the run-averaged transients graphically, or use piecewise approximation \cite{Jacobs:Purity2004}. 

The ideal protocol \textbf{I} requires ideal and instantaneous feedback which would be very difficult to achieve in practice. This motivated considering the no feedback case, for which the Bloch vector continually rotates about the $x$-axis.  The constant rotation takes the Bloch vector through the equatorial plane twice per cycle, this momentarily approximates the ideal feedback conditions.  This creates a minor increase in the purification rate, the dash-dot line (Fig.~\ref{fig:SpeedUp}).
 
When the feedback routine detailed in section~\ref{sec:sec4} is applied to the qubit with a best possible threshold value for $z_{\rm Limit}$, an almost optimum increase is achieved (dashed line) using a single practical control field. Hence we have shown that is is possible to create a control strategy to gain a significant amount of the ideal purification rate.

\begin{figure}[htbp]
	\centering
		\includegraphics[width=0.5\textwidth]{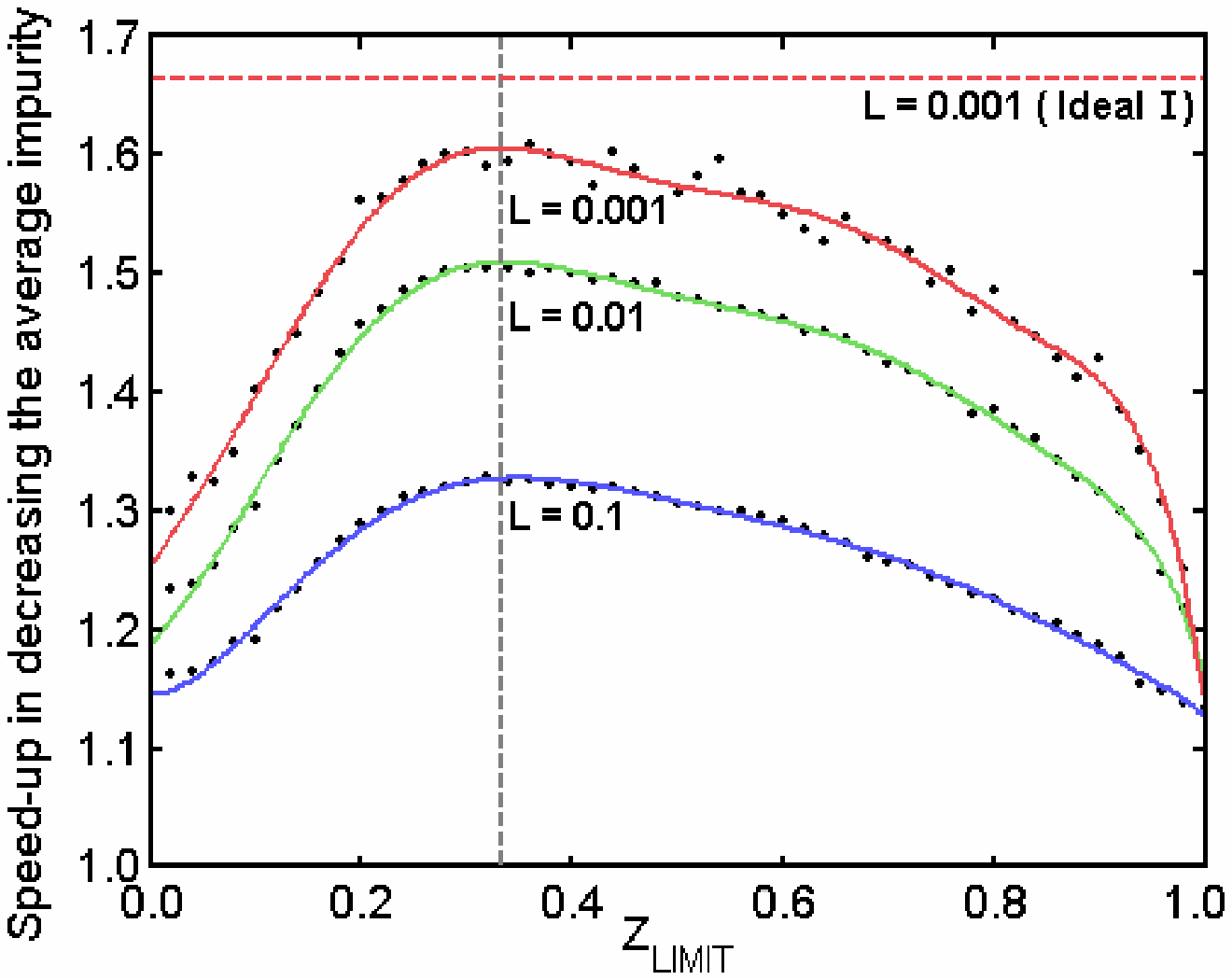}
	\caption{\label{fig:ZLimit} (Color online) This graph of the average purification rate improvement as a function of $z_{\rm Limit}$ shows the dependence on the thresholding at three different levels of purification.  Setting $z_{\rm Limit} \geqslant 1$ is equivalent to deactivating the feedback as the Bloch vector is constrained inside a unit sphere, alternatively using a small value threshold ($z_{\rm Limit}=0.1$) also yields lower performance due to the inherent measurement noise.  The horizontal dashed line is the  theoretical maximum performance increase given by Ideal Protocol \textbf{I} at $L = 0.001$.  The flattened peak for $L = 0.001$ near this `envelope' indicates a high tolerance of inaccuracies, with triggering between $z=0.2$ and $z=0.6$ being quite sufficient with an optimal value of $z_{\rm Limit} \approx 0.333$ as indicated by the vertical dashed line.}
	\vspace{0.25cm}
		\includegraphics[width=0.5\textwidth]{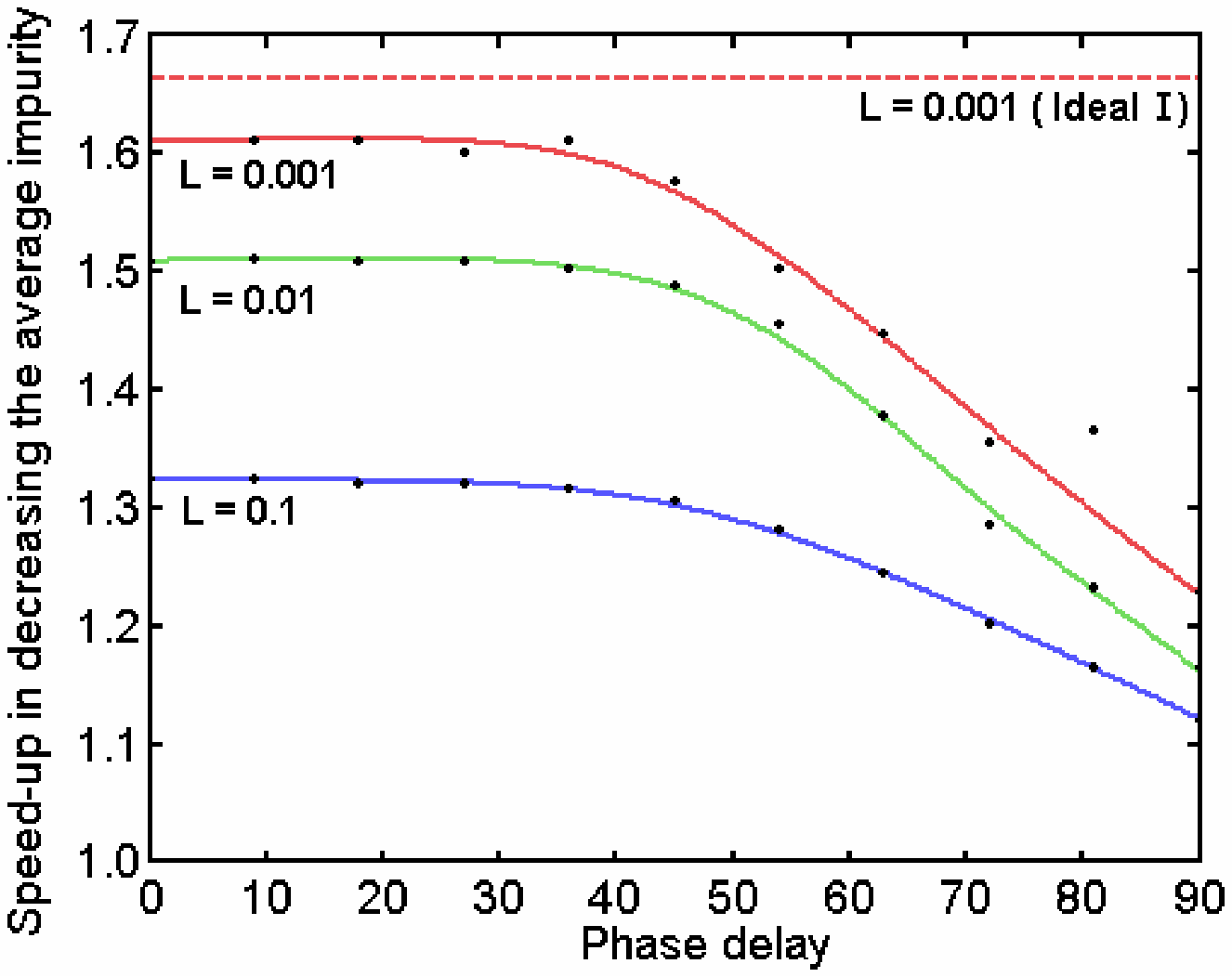}
	\caption{\label{fig:TimeDelay} (Color online) Using the optimal value for the thresholding, in this case $z_{\rm Limit} = 0.333$.  This graph of the average purification rate improvement as a function of the time delay starting from the top of the Bloch vector orbit (expressed as a phase delay with frequency $F_x$) suggests there is a reasonably large window of opportunity to apply the control fields and still achieve near optimal results.  Indeed this is a test of synchronisation, it could also be possible to allow the Bloch vector to make complete rotations and later apply the fields within this window.}
\end{figure}

The performance of the algorithm depends on the value of the $z_{\rm Limit}$ threshold, (Fig.~\ref{fig:ZLimit}).  We find good performance over a relatively large range, $0.2 < z_{\rm Limit} < 0.6$.  This implies the system can tolerate inaccurate and noisy thresholding, with only a minor penalty in the performance.  For low values of $z_{\rm Limit}$ the thresholding 
performs poorly due to the system operating under the `noise floor': due to the measurement noise during the control pulses triggering often recurs before the Bloch vector is returned to the vicinity of the $x$ axis.  For large values of $z_{\rm Limit}$ the trigger for applying the feedback is only activated at a late stage, and so the speed-up is severely reduced. 

Increasing the frequency $\nu$ of the Josephson junction aids the feedback procedure as the increased tangential velocity of the rotation reduces the time for the Bloch vector to reach the $x$-axis.  This is advantageous as the measurement noise disrupts the path taken (an example is provided in Figure~\ref{fig:ClampX}B), therefore the reduction in the time a point travels increases the probability of reaching close to the desired destination, the $x$-axis.  In addition, the relatively large Josephson junction energy gap reduces the possibility of thermal excitations between the two energy levels.

The anticipated disadvantage of using such high frequencies would be timing problems, although the control pulses should be feasible as the required $\pi$-pulses are often used in quantum information processing.  If time delays are a problem, then the qubit state can be allowed to rotate through several complete cycles if the radial growth of the Bloch vector is sufficiently small.  Figure~\ref{fig:TimeDelay} shows the effect of delaying the \textit{application} of the feedback as a phase angle from the top of the Bloch vector orbit, at the Josephson frequency of 10GHz for an optimum $z_{\rm Limit} \approx 0.333$.  Ideally, the feedback should be applied immediately ($0^\circ$) but it can be seen that there is an allowable delay with a small decrease in expected performance.

To summarize, we have shown in this section that the proposed method achieves a near optimum improvement in purification, and further improvement could be achieved  by an increase in Josephson junction frequency. It can be clearly observed in Figures~\ref{fig:ZLimit} and~\ref{fig:TimeDelay} that the method is quite tolerant of errors, and it is found to be sufficient to rotate the qubit back to the vicinity (rather than exactly on) the $x$ axis.

\section{\label{sec:sec6}FEEDBACK  PROTOCOLS II} 

In contrast to ideal protocol \textbf{I}, the ideal protocol \textbf{II} \cite{Wiseman:Purity} maximizes the stochastic terms in equations (\ref{eq:xyz}) by keeping the Bloch vector on the measurement axis. 

By using feedback to rotate the Bloch vector to the measurement axis, it has been shown recently that the majority of qubits reach a given level of purity faster than by using ideal protocol \textbf{I} \cite{Wiseman:Purity}.  It is the existence of rare but extremely long purification times which makes the average purification time longer for this scheme than that for the deterministic ideal protocol \textbf{I}.

For a qubit which has either Hamiltonian evolution around the measurement axis or no Hamiltonian at all, this protocol requires no controls to implement., as is not a problem and the Bloch vector naturally diffuses along the measurement axis between the two possible outcomes (the poles).  However, there is an issue when a Hamiltonian takes the Bloch vector away from the measurement axis, meaning the effect of the system measurement will be less than ideal.  Unfortunately, this is the case for our superconducting charge qubit, where the continual $x$-axis rotation takes the Bloch vector away from the $z$-axis and passes through the $x$-$y$ plane before returning to the $z$-axis (Fig.~\ref{fig:Spiral}).  This helps for the purpose of maximizing the rate of purity increase of a single qubit, but is harmful for the purpose of minimizing the average time for a qubit  to reach a given purity.

\subsection{\label{sec:sec6pA} Ideal Protocol II}

Assuming ideal and instantaneous controls, it would be possible to apply feedback to rotate back perfectly to the $z$-axis after each measurement.  If this were possible the purity would evolve as per equation (\ref{eq:purity_worse}).

\subsection{\label{sec:sec6pB} Flux-controlled Hamiltonian Feedback}

A tunable flux-controlled Josephson junction can be used to change the $x$-axis rotational frequency, slowing the passage of the Bloch vector near the poles and speeding it through the equatorial plane~\cite{Ralph:Aero}.  This is the dual of the method described in section \ref{sec:sec4pB}.

\subsection{\label{sec:sec6pC} Practical Protocol II}

\begin{figure}[htbp]
	\centering
		\includegraphics[clip,width=0.4\textwidth]{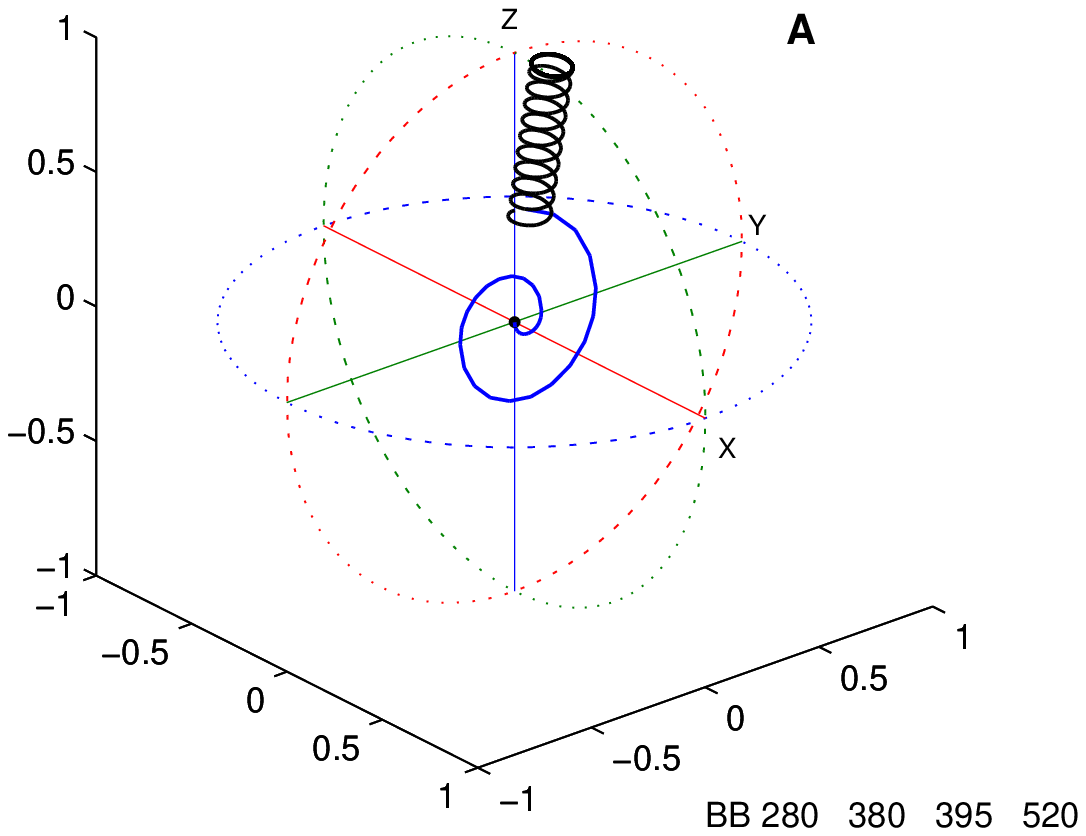}
		\vspace{0.5cm}
		\includegraphics[clip,width=0.4\textwidth]{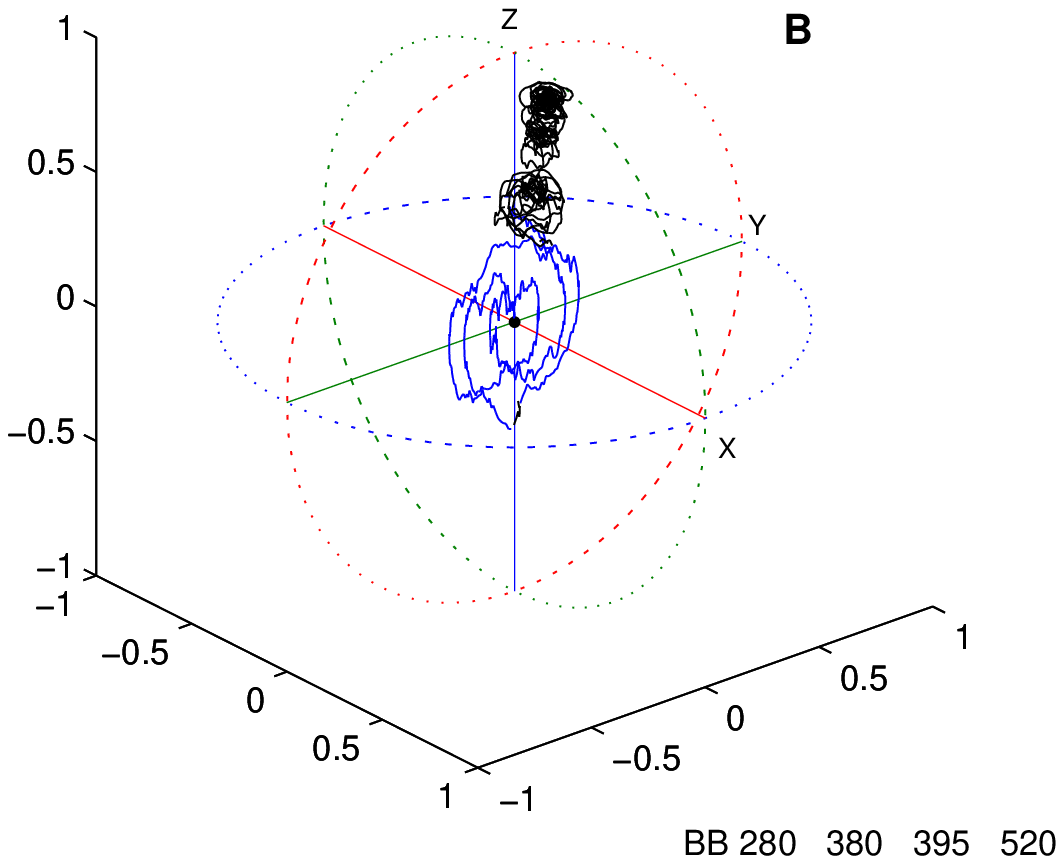}
	\caption{\label{fig:ClampZ} (Color online) (\textbf{A}) Pictorial representation showing the intended path of the Bloch vector.  The feedback control is initially off, so that the Bloch vector continually rotates and grows around the $x$-axis, this is to allow the peak value of $z$ to become more distinct.  Once the Bloch vector has exceeded a threshold and is at a maximum or minimum, the high frequency $z$-rotations are applied which locks the Bloch vector to the measurement axis.	(\textbf{B}) An example trajectory showing the effect of measurement noise on the path of the Bloch vector.  Whilst the noise is significant, the overall shape of the stochastic trajectory is similar to the conceptual path shown above.  The shape confirms that the Bloch vector is confined to the vicinity of the measurement axis ($z$-axis) as required.}
\end{figure}

It is anticipated that there would be difficulties in adapting the previously described method, for rotating to the $x$-axis, to  rotate to the $z$-axis instead.  Due to the constant $x$-axis rotations and the switchable $z$-axis rotations, the nature of the problem is not symmetrical.  Whenever the $z$-axis rotations are removed, the Bloch vector will still rotate about the $x$-axis therefore taking the vector away from the required $z$-axis.  Unless the experimental apparatus can measure, process and apply a correcting control field within a fraction of a cycle, the application of feedback will be futile as a complete cycle about the $x$-axis will have been made anyhow.

To solve this, we propose a very simple feedback protocol where the Bloch vector rotates about a tilted axis almost parallel to the measurement axis (Fig.~\ref{fig:ClampZ}A).  

\begin{figure}[htbp]
	\centering
		\includegraphics[width=0.5\textwidth]{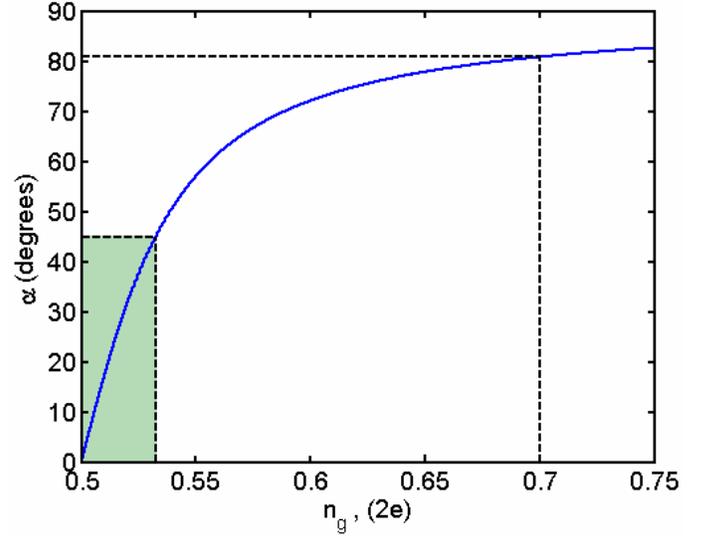}
	\caption{\label{fig:alpha} (Color online) Angle $\alpha$ as a function of bias, $n_g$.  This illustrates the operating range of the two techniques described in this paper.  The green region indicates the necessary angles and biases for practical protocol \textbf{I}, the range of which is intrinsically constrained by the maximum value of $\alpha = 45^\circ$, $(0^\circ \leq \alpha \leq 45^\circ)$.  The solitary dashed line illustrates a possible bias value for generating the tilted axis of practical protocol \textbf{II}. Notice that only one value is required.  We require $\alpha$ to be as close to $90^\circ$ as possible, however if the bias is set too far from 0.5, there is a risk of accessing an unwanted third state.  For large bias values the gains in $\alpha$ are actually quite minimal.}
\end{figure}

Initially, no $z$ control is applied and the Bloch vector is allowed to rotate about the $x$-axis.  The effect of the weak measurement is to pull the Bloch vector towards the poles, and a spiral growth results.  This initial period allows an experimentalist to detect a sizeable peak or trough in the $z$ measurement record corresponding to the phase of the oscillation in $z$, indicating when the Bloch vector is near to the $z$-axis.  When at this point, if the threshold value $(z_{\rm Limit} = 0.333)$ is exceeded, the strong $z$ control is applied, creating a rotation about an axis that should be almost parallel to the $z$-axis.  If the initial detection is completed successfully, the Bloch vector should be near the $z$-axis and will now travel in a tight spiral close to the $z$-axis (Fig.~\ref{fig:ClampZ}A).  Alternatively, if the Bloch vector was somewhere near the $y$-axis, for example due to an initial delay, the orbital path about this tilted axis would be much wider and fail to coincide with the actual $z$-axis.  One could instead simply apply this $\sigma_z$ control from the start of the purification process, to ensure that the qubit always remains close to the $z$-axis. However, there is little benefit in applying the control at the early stages because the performance gain close to the centre of the Bloch sphere is small. 

The definition of $\alpha$ as the angle of the axis of rotation between the $x$ and $z$ axes, still holds for this method.  Equation (\ref{eq:alpha_freq}) defines $\alpha$ as a function of Bloch sphere coordinates. Here we express it in terms of the system frequencies: the constant Josephson junction frequency $\omega_x$ and the bias control frequency $\omega_z$:
\begin{eqnarray}
	\alpha & = & \frac{1}{2}\sin^{-1}\left(\frac{\omega_z}{\omega_x}\right).
	\label{eq:alpha_freq}
\end{eqnarray}

The magnitude of the bias field should not be too large or the next charge state may be accessed and the two state approximation would be violated.  Figure~\ref{fig:alpha} shows how $\alpha$ varies as a function of the bias control, plotted until $n_g = 0.75$ as this is halfway between the charge states.  A bias value of $n_g = 0.70$ is chosen in the simulations below to reduce the possibility of accessing a new state. This gives $\alpha = 82^\circ$ (see Fig.~\ref{fig:alpha}). Increasing $n_g$ gives minimal  gain in $\alpha$ as the angle asymptotes to $90^\circ$.   Ideally we would have $\alpha = 90^\circ$ but we find an angle of  $82^\circ$ gives acceptable performance.

\section{\label{sec:sec7}RESULTS II} 

\begin{figure}[htbp]
	\centering
		\includegraphics[width=0.5\textwidth]{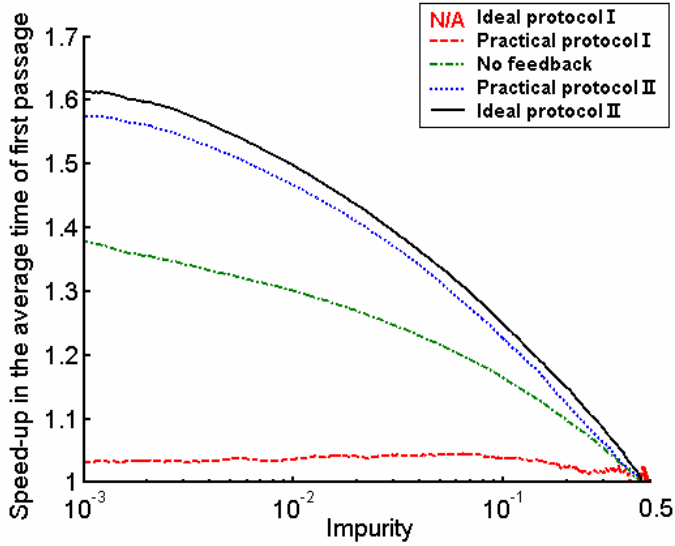}
	\caption{\label{fig:TimeFP} (Color online) The graph of the improvement, in comparison with ideal protocol \textbf{I}, in achieving the time of first passage shows that the practical protocol proposed in this section (blue line) achieves similar performance to that of the ideal protocol \textbf{II} (black line), which required perfect instantaneous feedback.  In addition, the practical protocol described in section~\ref{sec:sec4pC} (red line) has little performance gain, as expected.  It should also be noted that using no feedback in this system still yields a minor improvement (green line) over having no Hamiltonian evolution at all}
\end{figure}

Figure~\ref{fig:TimeFP} shows that the performance gain for the practical protocol II is near that of ideal protocol II. This indicates the desired operation: a practical implementation of rotating to the $z$-axis despite the presence of the constant $x$-axis rotations due to the qubit tunnelling.  Thus considering Figs.~\ref{fig:SpeedUp}  and \ref{fig:TimeFP}, we see that both the practical protocols \textbf{I} and \textbf{II}, emulate the ideal counterparts. 

\begin{figure}[htbp]
	\centering
		\includegraphics[width=0.50\textwidth]{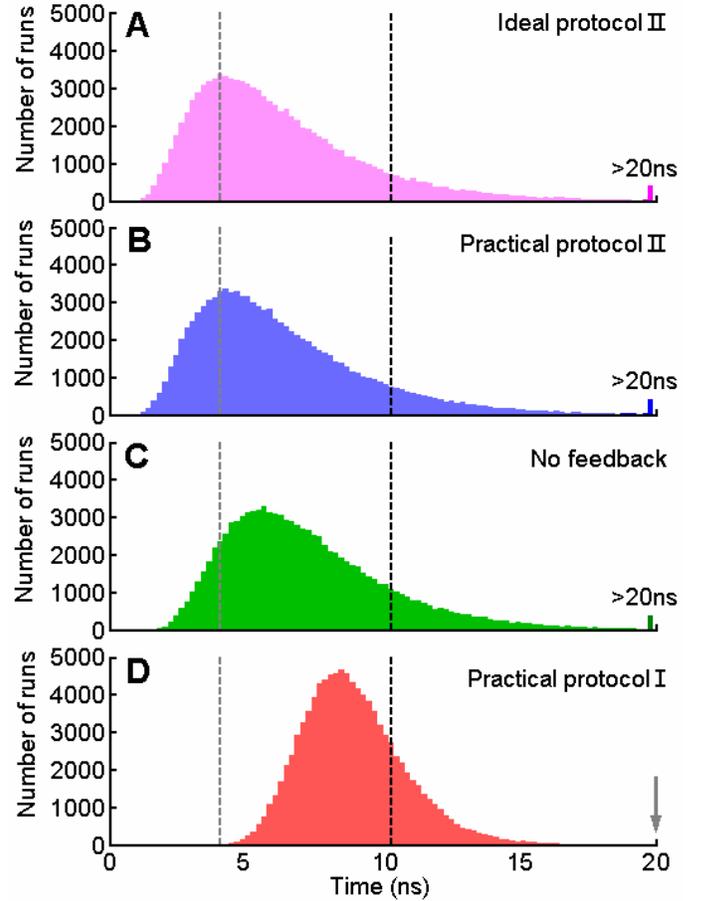}
	\caption{\label{fig:TAP} (Color online) Plotting the distribution of times to reach a given impurity of $1\times10^{-3}$ clearly shows that the majority of simulated runs (100,000 runs) of the other protocols reach the target impurity before the deterministic ideal protocol \textbf{I}, (indicated by right most dotted line), which greatly reduces the \textit{average impurity}.  Due to the deterministic nature of ideal protocol \textbf{I}, it guarantees that all qubits reach a set impurity in a given time.  However it has been suggested that the poor average performance of rotating to the $z$-axis is due to the existence of a few important and extraordinarily long duration runs ($\geq20$ns simulation time) which have a significant effect on the calculation of the average.}
\end{figure}

\begin{figure}[htbp]
	\centering
		\includegraphics[width=0.50\textwidth]{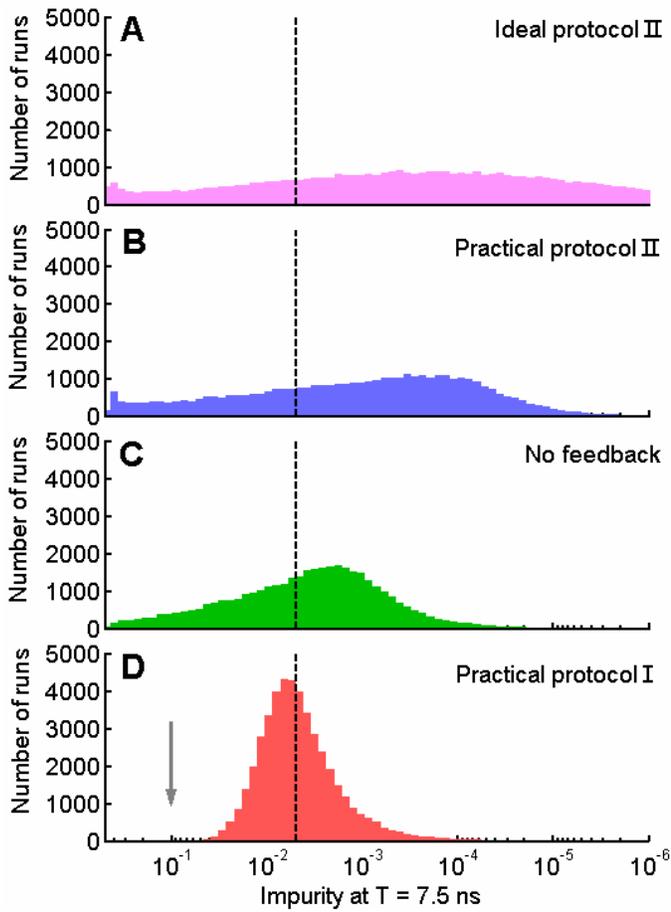}
	\caption{\label{fig:PAT} (Color online) The histograms (50,000 runs) of the final impurity at the simulation end time of 7.5ns, demonstrates the large range of final impurities given ideal protocol \textbf{II} (A), allowing access to better purity in a shorter time at the risk of worse purity on some occasions.  Interestingly, when perfect feedback is replaced with practical protocol \textbf{II} (B), the range of very small impurities (below $10^{-5}$) is not accessible.  In addition, when using practical protocol \textbf{I} we can (for these example values) be confident that by 7.5ns almost all impurities will be at least smaller that $5\times10^{-2}$, indicated by the arrow.}
\end{figure}

To demonstrate the usefulness of rotating the Bloch vector to the $z$-axis, we examine the statistical distributions for the five options described in this paper. In Figs.~\ref{fig:TAP} and \ref{fig:PAT} the following notation is used: \textbf{A} -- Perfect rotations to the $z$-axis (Ideal protocol \textbf{II}  \cite{Wiseman:Purity}), \textbf{B} -- Rotations about an axis almost parallel to the $z$-axis (Practical protocol \textbf{II}), \textbf{C} -- $\sigma_x$ tunnelling Hamiltonian only (No feedback), \textbf{D} -- Rotating to the $x$-axis (Practical protocol \textbf{I}), and finally, \textit{black dashed line} -- Perfect rotations to $x$-$y$ plane (Ideal protocol \textbf{I} \cite{Jacobs:Purity2003}).   The component values and measurement strength used to simulate the following histograms can be found in Appendix \ref{sec:appA}.

Figure~\ref{fig:TAP} depicts the distributions of the times of first passage, that is, the time for the stochastic impurity to fall below a given impurity of $1\times10^{-3}$.  Each histogram comprises  $10^5$ independent simulation runs separated into 100 bins.  Figure~\ref{fig:TAP}A confirms that for ideal protocol \textbf{II}, the majority of runs reach the required purity at times earlier than that predicted by ideal protocol \textbf{I}. The modal value for ideal protocol {\bf II} (grey dashed line in all the plots) is 4.0ns as opposed to 10.4ns for ideal protocol \textbf{I}.  It is apparent that the poor \textit{average} performance is due to the existence of a small number of extraordinarily long duration runs which have a significant effect on the average purity.  Note that the many runs which do not reach the required purity by the end of the simulation ($T_{\rm max} = 20$ns) and are included in the histogram marked as `$>20$ns'.

Figure~\ref{fig:TAP}B shows the remarkable similarity between the ideal and practical protocol \textbf{II} distributions, with the modal values in close alignment.  This would indicate that rotating about a tilted $z$-axis is a near optimal approach for rotating to the $z$-axis, given the presence of a constant $x$ or $y$ rotation.

When no feedback is applied, there is a constant rotation about the $x$-axis.  This rotation momentarily passes the Bloch vector through the $z$-axis and the $x$-$y$ plane, creating a compromise between the two ideal purification methods (Fig.~\ref{fig:TAP}C).

Rotating the Bloch vector to the $x$-axis, (practical protocol \textbf{I} as described in section~\ref{sec:sec4pC}), yields a distribution of smaller variance. Indeed it should be noted, as indicated by the arrow, no runs exceeded the maximum simulation time of $20$ns.  The modal value is closer to that of ideal protocol {\bf I} ($10.4$ns). The non-zero variance is due to the need to allow the Bloch vector to grow and rotate about the $x$-axis; the non-zero $z$ component (Fig~\ref{fig:ClampX}B) contributes measurement noise to the purity [see \erf{eq:xyz}].

Figure~\ref{fig:PAT} shows the distribution of impurities at $t=7.5$ns, which is the time at which the impurity under ideal protocol \textbf{I} reaches $5\times10^{-3}$.  Each histogram contains 50,000 runs separated into 50 logarithmically spaced bins. 

Comparing Figs.~\ref{fig:PAT}A and~\ref{fig:PAT}D we see a  dramatic difference in the spread of values by many orders of magnitude. The deterministic natures of both ideal protocol \textbf{I} and the reduced stochasticity of the more practical protocol \textbf{I} can easily be observed. Of particular interest is the area corresponding to high impurity indicated by the arrow.  In Figure~\ref{fig:PAT}D this region is mostly unoccupied, but the other three histograms which do not employ protocol \textbf{I} have high occupancy.  This implies that although these three methods can potentially reach very low impurities, it is done at the risk of ending with a high impurity. 
  
Interestingly, Figure~\ref{fig:PAT}B follows a similar profile to Figure~\ref{fig:PAT}A until the impurity is of the order $10^{-5}$, when smaller impurities become inaccessible.  This is due to a \textit{mushrooming} effect which creates an end cap to the expected path of the Bloch vector. The end cap occurs whenever the Bloch vector is near a pole at the surface of the Bloch sphere and is due to the weak measurement noise. This can be further explained by examining Eqs.~(\ref{eq:xyz}) for weak measurement in the Bloch sphere representation when the Bloch vector is near a $z$-axis pole.  It can be seen that as $|z|$ approaches one, the random contribution of the Weiner increment becomes much larger.  As the $x$ and $y$ values are non-zero (due to the off-axis rotations removing the Bloch vector from the $z$-axis), the `large' random changes in $x$ and $y$ combined with the constant rotation due to Hamiltonian evolution makes it naturally improbable that the Bloch vector will settle exactly on the $z$-axis.  Hence, in practice access to the smallest impurities may be difficult without increasing the measurement strength $\gamma$ relative to $\nu$.

\section{\label{sec:sec8} CONCLUSIONS} 

We have considered two techniques for rapid state purification for use with a model superconducting charge qubit with a single control field and continuous weak measurement.  We show that near optimal results can be obtained using a realistic implementation of feedback control.  For practical protocol \textbf{I}, the feedback is simple to calculate and uses constant amplitude $\pi$-pulses that are applied for time scales that are comparable with the natural period of the qubit evolution.  In addition, as the maximum angle between the rotation axis and the $x$-$y$ plane ($\alpha$) is $45^\circ$, the range of bias controls is confined to a small range of values close to the default bias condition. For practical protocol \textbf{II}, the feedback control is simply maintained, once triggered.

If an experimenter wished to ensure that the majority of qubits reach the same level of purity at the same time, ideal protocol \textbf{I} or the more practical implementation of rotating to the $x$-axis as described in section~\ref{sec:sec4pC} should be used (practical protocol \textbf{I}).  Alternatively, if the objective is to maximise the number of qubits attaining a given level of purity, the experimenter should choose practical protocol \textbf{II} described in section~\ref{sec:sec6pC} and based on ideal protocol \textbf{I}. As the techniques need only keep the Bloch vector close to the ideal conditions, the practical protocols are expected to be robust to a variety of control errors (e.g. time delays, magnitude of bias, pulse duration).  

Both of the practical protocols (\textbf{I} and \textbf{II}) described in this paper operate with continuous rotations about the $x$-axis, which are generated by the constant tunnelling frequency of the Josephson junction (which is the realistic scenario for practical superconducting charge qubits).  Numerical simulations demonstrate that both of the practical protocols perform well and are not adversely affected by this constraint on the controls allowed in the system. In fact, the constant $\sigma_x$ term arising from the tunnelling is essential to the correct operation of practical protocol \textbf{I}. If these continuous rotations are either naturally occurring or can be applied, it opens the possibility of implementing these purification techniques in other systems that contain such Hamiltonian evolution.

\begin{appendix}

\section{\label{sec:appA} Table of values}

\noindent Values are constant and consistent for all simulations, in line with experimental values quoted in reference~\cite{Pashkin:Nature}.

\begin{table} [h]
	\caption{\label{tab:components}Components}
	\begin{ruledtabular}
		\begin{tabular}{clc}    
			 & Description & Typ.\\
			\hline
			\\
			$\nu/2\pi$ 	& Josephson junction energy 					& 10GHz\\
			$C_J$ 			& Josephson junction capacitance 			& 500aF\\
			$C_g$ 			& Qubit-Grounded Bulk capacitance			& 0.5aF\\
			$C_p$ 			& Electrodes parasitic capacitance 		& 1.0aF\\
			\\
			\hline
			\\
			$\gamma$		& Measurement strength constant				& $75\times10^6$\\
			\\
		\end{tabular}
	\end{ruledtabular}
\end{table}

\section{\label{sec:appB} Weak measurement in the Bloch sphere representation}

Here we state the Cartesian equations for the random incremental changes in $x$, $y$ and $z$ for each time step $dt$ due to a continual weak measurement process with measurement strength $\gamma$.  These equations can be found in reference \cite{Jacobs:Purity2004}, and they can be derived by working through equation (\ref{eq:WeakRho}) using the Pauli matrix identities and equating the resulting density matrix elements with the Bloch vector coordinate equations: $x = Tr\left\{\rho_c\sigma_x\right\}$, $y = Tr\left\{\rho_c\sigma_y\right\}$ and $z = Tr\left\{\rho_c\sigma_z\right\}$.

\begin{subequations}
  \begin{eqnarray}
	  dx  & = & -(4\gamma dt + z\sqrt{8\gamma}dW)x
	  \label{eq:x}
	  \\
	  dy  & = & -(4\gamma dt + z\sqrt{8\gamma}dW)y
	  \label{eq:y}
	  \\
	  dz  & = & (1-z^2)\sqrt{8\gamma}dW
	  \label{eq:z}
	  \\
	  L & = & \frac{1}{2} \left( 1 - x^2 - y^2 - z^2 \right)
	  \label{eq:P}
	  \end{eqnarray}
	\label{eq:xyz}
\end{subequations}

\noindent This set of simultaneous stochastic differential equations is not trivial to solve.  Indeed, equation (\ref{eq:purity_worse}) is the special case where $x = y = 0$, and yet yields an integral that appears to have no analytical solution \cite{Jacobs:Purity2003}.

\end{appendix}

\begin{center}
	\textit{Acknowledgments}
\end{center}

\noindent EJG would like to acknowledge the support of the Department of Electrical Engineering and Electronics, and a University of Liverpool research scholarship.  CH and JFR would like to acknowledge the support of an ESPRC grant: EP/C012674/1.  HMW was supported by the Australian Research Council and the State of Queensland.  KJ would like to acknowledge the support of the The Hearne Institute for Theoretical Physics, The National Security Agency, The Army Research Office and The Disruptive Technologies Office.

\bibliography{PurityBib_a}

\clearpage

\end{document}